\documentstyle[preprint,eqsecnum,aps]{revtex}

\begin{document}
\draft
\preprint{Univ. of Waterloo}
\title{Revisiting Black-Scholes Equation 
}
\author{D. F. Wang$^\dagger$}
\address{University of Waterloo}
\date{April, 1998}
\maketitle
\begin{abstract}

In common finance literature, Black-Scholes partial 
differential equation of option pricing 
is usually derived with no-arbitrage
principle. Considering an asset market, Merton applied  
the Hamilton-Jacobi-Bellman techniques of his continuous-time
consumption-portfolio problem,
deriving general equilibrium relationships among the securities
in the asset market. In special case where the interest
rate is constant, he rederived the Black-Scholes partial differential
equation from the general equilibrium asset market. In this work, 
I follow Cox-Ingersoll-Ross formulation to consider an economy
which includes  
(1) uncertain production processes, and (2) the random
technology change. Assuming a random production
stochastic process of constant drift and
variance, and assuming a random technology change to follow a log
normal process, the equilibrium point of this economy will
lead to the Black-Scholes partial differential equation
for option pricing.  
\end{abstract}
\pacs{PACS number: }


\section{Introduction}

There are two important concepts in theories of asset pricing in 
modern finance.
The first is the principle of no arbitrage, and the second one
is the idea of general equilibrium\cite{boyle,pliska,merton1}.  
In financial world, one is 
not expected to gain a riskless free lunch out of nothing.
This is the basic idea of no arbitrage. Mathematically, in 
a complete financial market, no arbitrage implies the existence
of a unique risk neutral probability measure $Q$, and vice versa. 
In this system,
any contingent claim discounted by the bank account process will 
form a martingale in the probability space of risk neutral measure
$(\Omega, F, Q)$, that is, $X_t/B_t=E^Q(X_T/B_T|F_t)$, with $t<T$
and $F_t$ as filtration. The principle of no-arbitrage has considerable
applications in asset pricing, and many useful asset pricing models
are derived using this idea. Within this approach, one does not have
to know the investor's risk preferences explicitly. Intuitively,
one can understand why this principle should hold in pricing assets.
Imagine a financial world in which the prices of some assets are such that
some riskless gain can be obtained by investors of zero initial wealth. 
Investors will rush to the position to take advantage of this arbitrage
opportunity, and a dynamic process in which the asset prices will
adjust themselves will take place, 
so that the arbitrage opportunity no longer exists.
In other words, the asset prices are determined by the condition that  
no arbitrage exists. This principle is in analogy to the physics idea 
that no machine can produce non-zero energy out of nothing.  
 
Apart from this no-arbitrage concept, the second important approach is  
application of the concept of general equilibrium. Within the framework, 
we consider an economy of homogeneous individuals, so that a representative
agent's economical behavior is studied. Contingent claims are
sold and bought by the representative agent in the economy. 
When the equilibrium is attained,
the representative agent should maximize his/her expected utility function,
and he/she will have no more intention to trade contingent claims.
Such equilibrium condition will determine the prices of the contingent claims
in terms of the fundamental parameters which specify the economy.  
Theoretically, this approach relates the asset prices to the fundamentals of 
the economy in which the agent is living. In academic sense, 
this appears to be an approach more fundamental
than the first one of applying no-arbitrage argument. 
Within the framework of general equilibrium, one starts with 
the risk preferences of the representative agent. 

Besides considerable numerical efforts in asset pricing,
such as the well-known MC method of pricing derivatives pioneered by Boyle\cite{boyle2}, 
theoretical and analytical studies for closed form of pricing 
formula of derivatives have been of considerable interests to 
both academic researchers and practioners.
The option pricing model of Black-Scholes has been very popular, since
it was developed by Black and Scholes\cite{black}. Computationally, the model is simple  
to handle, and the closed forms of European call and put options
can be obtained explicitly.    
In common financial literature, the Black-Schole partial differential 
equation is derived
using the concept of no-arbitrage\cite{book}. By introducing a continuously
rebalanced risk-free portfolio consisting of an option and underlying
stocks, the return of such a portfolio should match the return on
risk-free bonds, in absence of arbitrage. This will give rise to
the partial differential equation satisfied by the option price\cite{self}.
An alternative way of using no-arbitrage principle is to replicate
an option's return stream by continuously rebalancing a self-financing
portfolio containing stocks and bonds.  
 
It is nature to ask whether one can derive the Black-Scholes partial
differential equation, instead of using the previous no-arbitrage
approach, but using the principle of general equilibrium.
Merton applied the Hamilton-Jacobi-Bellman technique of 
his continuous-time consumption-portfolio problem\cite{merton3,merton4} to an asset market,
and he derived general equilibrium relationships among the securities
in the asset market\cite{merton2}. In the special case where the interest
rate is constant, he rederived the Black-Scholes partial differential
equation from the equilibrium asset market\cite{merton2}.  
In this work,
I follow Cox-Ingersoll-Ross formulation\cite{cox1,cox2} to consider an economy
which includes
(1) uncertain production processes, and (2) random
technology change. Assuming a random production
stochastic process of constant drift and
variance, and for the random technology change following a log
normal process, it is shown that the equilibrium point of this economy will
lead to the Black-Scholes partial differential equation
for option pricing.

\section{General Equilibrium}

In this section, we use the formalism of the works of Cox, Ingersoll 
and Ross\cite{cox1,cox2}. They considered a general equilibrium problem of 
an agent investing and consuming in a continuous time fashion,
where uncertain production and a random technology change
are taken into account. 
In a similar way as in Merton's problem\cite{merton1,merton2,merton3,merton4}, 
they derived the optimal consumption-investment control equation
with Hamilton-Jacobi-Bellman technique. Assume that $a_i(s) W(s)$ 
is the investment in the i-th production plan, and $b_i(s) W(s)$ the 
investment in the i-th contingent claim. At the general equilibrium, 
the market clearing conditions require that $b_i=0$ for all $i$, and 
$\sum_{i} a_i = 1$\cite{cox1,cox2}. CIR used this general equilibrium concept to propose
their well-known interest rate model. The price of the 
corresponding zero-coupon bond is shown to satisfy a partial differential
equation. 

On the other hand, we know that within the framework of Black-Scholes,
a deterministic interest rate $r$ is used in the derivation based on
no-arbitrage principle. Therefore, our idea is to find what type of 
uncertain production process and random technology change
will give rise to an economy  
whose general equilibrium point is a constant interest rate. For such economy,
the price of any contingent claim will satisfy the Black-Schole partial
differential equation. 

Let us first review the work of Cox, Ingersoll and Ross and
we follow the notations used by them\cite{cox1,cox2}. Suppose that the 
utility function of the representative agent is additive. Consider the 
following value function:
\begin{equation}
K(v, W(t), Y(t), t) = E[\int_t^T U(v(s), Y(s), s) ds],
\end{equation}
where $v(t)$ is an admissible feedback control, and the expectation
is conditional on that information before or equal to time $t$ is known 
to the investor. $U$ is the van-Neumann-Morganstern utility function.
The control is $v=(a(s)W(s), b(s)W(s), C(s))$ with $C(s)$ as the 
consumption rate.  $Y$ is the state variable representing the uncertain  
technology. The indirect utility function $J(W, Y, t)$ is 
$max K(v, W(t), Y(t), t)$ among all possible controls $v\in V$, as 
defined in the works of CIR. The indirect utility function can be easily 
shown to be an increasing and concave function of the current wealth.  

In this economy, the production process of the system is defined to be 
\begin{equation}
d\eta(t)=I_\eta \alpha(Y,t) dt + I_\eta G(Y, t) d\omega(t),
\end{equation}
and random technology change $Y$ takes the following form
\begin{equation}
dY(t)=\mu(Y,t)dt+S(Y,t)d\omega(t),
\end{equation}
where $\omega(t)$ is a standard $(n+k)$-dimensional Brownian motion.
Given the control $(aW,bW,C)$, the budget constrain of the agent is 
given by
\begin{equation}
dW=[\sum_{i=1}^n a_i W(\alpha_i-r)+ \sum_{i=1}^k b_i W(\beta_i-r)+
rW-C] dt+
\sum_{i=1}^n a_i W\sum_{j=1}^{n+k} (g_{ij} d\omega_j)+W \sum_{i=1}^k b_i
(\sum_{j=1}^{n+k} h_{ij} d\omega_j),
\label{eq:budget} 
\end{equation}
The optimal control is given by the 
Hamilton-Jacobi-Bellman equation, and we therefore have the following
first order conditions:
\begin{eqnarray}
&&\Psi_C=U_C-J_W\le 0,\\
&&C\Psi_C=0,\\
&&\Psi_a=[\alpha -r 1] WJ_W +[GG'a+GH'b]W^2J_{WW}+GS'WJ_{WY}\le 0\\
&&a'\Psi_a=0,\\
&&\Psi_b=[\beta-r1]WJ_W+[HG'a+HH'b]W^2J_{WW}+HS'WJ_{WY}=0,
\end{eqnarray}
where the subscripts denote partial derivatives. It is important to
note that we must have $C(t)\ge 0$ and $W(t) \ge 0$ to satisfy 
feasibility conditions\cite{huang,merton1}.  

The market clearing conditions require 
that $\sum_i a_i(s)=1$ and $b_i(s)=0, \forall i $
for the system. These conditions combined with the above
optimal control equations will enable us to find the corresponding
production investment $a^\star$ and the optimal consumption $C^\star$.
The interest rate, as a function of the wealth
$W(s)$, the state variable $Y(s)$ and time $s$, was found to be:
\begin{equation}
r(W,Y,t)=(a^\star)' \alpha -({-J_{WW}\over J_W})({varW\over W})
-\sum_{i=1}^k {-J_{WY_i}\over J_W} {Cov(W,Y_i)\over W}.
\end{equation}
At the point of general equilibrium, the price of any contingent claim
$F(W, Y, t)$ should satisfy the following partial differential equation
\begin{eqnarray}
&&{1\over 2}(varW)F_{WW}+\sum_{i=1}^k Cov(W,Y_i)F_{WY_i}
+{1\over 2}\sum_{i=1}^k\sum_{j=1}^k Cov(Y_i,Y_j)F_{Y_iY_j}
+[r(W,Y,t)W-C^\star(W,Y,t)]F_W\nonumber\\
&&+\sum_{i=1}^k F_{Y_i}[\mu_i-({-J_{WW}\over J_W} Cov(W,Y_i))
-\sum_{j=1}^k
{-J_{WY_i}\over J_W} Cov(Y_i,Y_j)]
+F_t-r(W,Y,t)F+\delta(W,Y,t)=0.
\label{eq:value1} 
\end{eqnarray}
This last equation is the Theorem 3 proved in the paper of Cox-Ingersoll-Ross\cite{cox1}.
It is the fundamental valuation equation for the contingent claims. 

Now, let us assume that the risk preference of the consumer-investor
is Bernoulli logarithmic utility function, 
$U(C(s),Y(s),s)= e^{-s\rho} lnC(s)$, with some time discount prefactor. 
This risk preference was 
discussed in previous works\cite{merton3,merton4}. 
In the following, I use the scaling argument
of physics-type to rederive some functional form of the indirect 
utility function, consumption rate, and so on\cite{wang}. 
Consider a scaling transformation defined below:
\begin{eqnarray}
&&W(s)\rightarrow A W(s)\nonumber\\
&&C(s)\rightarrow A C(s)\nonumber\\
&&a(s)\rightarrow a(s)\nonumber\\
&&b(s)\rightarrow b(s)\nonumber\\
&&r(s)\rightarrow r(s)\nonumber\\
&&Y(s)\rightarrow Y(s)\nonumber\\
&&\eta(s)\rightarrow \eta(s),
\label{eq:scaling1} 
\end{eqnarray} 
where $A$ is any positive constant, and $t\le s \le T$.
Under this scaling transformation, the budget constrain 
Eq.~(\ref{eq:budget}) will remain unchanged. The value function
$K(v,W,Y,t)$ will scale in the following way:
\begin{equation}
K(v,W,Y,t)\rightarrow {1\over\rho} [e^{-\rho t} -e^{-\rho T}] lnA +K(v,W,Y,t), 
\label{scaling2}
\end{equation}
which can be verified easily. Therefore, at the optimal control, we shall
have the scaling laws for the indirect utility function, the optimal
investment and consumption rate as below:
\begin{eqnarray}
&&J(AW, Y, t)={1\over \rho} (e^{-\rho t}-e^{-\rho T}) lnA + J(W,Y,t),\nonumber\\
&&C^\star(AW,Y,t)=A C^\star(W,Y,t),\nonumber\\
&&a^\star(AW,Y,t)=a^\star(W,Y,t),
\end{eqnarray}
which holds for any positive constant $A$. Since these scaling laws hold
for any positive constant $A$, we can solve these equations to find
the functional dependence on the wealth $W$. 
It is straightforward to find the following relationships:
\begin{eqnarray}
&&J(W,Y,t)={1\over \rho}[e^{-t\rho}-e^{-T\rho}]lnW +f_1(Y,t),\nonumber\\
&&C^\star(W,Y,t)=f_2(Y,t)W,\nonumber\\
&&a^\star(W,Y,t)= f_3(Y,t),
\label{eq:law}  
\end{eqnarray}
where the functions $f_1(Y,t),f_2(Y,t)$ and $f_3(Y,t)$ are solely dependent
on the variable $Y$ and time $t$, and $f_1(Y,T)=0$.  
These relations hold for the consumer-investor 
with the Bernoulli logarithmic utility function, and for
any random technology change and random production processes. 
In this case, we have $J_{WW}W/(J_W)=-1$ and $J_{WY}=0$. 

For this consumer-investor of Bernoulli logarithmic utility function,
we further assume that the random production processes have 
constant drift and constant variance:
\begin{eqnarray}
d\eta(t)&&=I_\eta \alpha(Y,t) dt + I_\eta G(Y, t) d\omega(t)\nonumber\\
&&=I_\eta \alpha dt + I_\eta G d\omega(t),
\end{eqnarray}
where both $\alpha$ and $G$ are time independent constants.
With the first order optimal conditions, market clearing conditions,
and using the scaling properties of the indirect utility functions, 
one can derive the investment rate
\begin{equation}
a^\star=(GG')^{-1} \alpha +{[1-1'(GG')\alpha]\over 1'(GG')1 } (GG'),
\end{equation}
which is also a time independent constant. The interest rate $r(W,Y,t)$ 
is given as 
\begin{equation}
r(W,Y,t)=(a^\star)'\alpha +(a^\star)'(GG')a^\star(-1),
\end{equation}
which is a time independent constant. Therefore, we see that for 
a random production process of constant drift and constant variance,
the consumer-investor of Bernoulli logarithmic utility function
will have a constant interest rate $r$ when the general equilibrium is
reached, consistent with our intuitive expectation.

Furthermore, we assume that the state variable $Y$ describing
random technological change is one dimensional, and 
it follows a stochastic process given by
\begin{eqnarray}
dY=&&\mu(Y,t)dt+S(Y,t)d\omega(t)\nonumber\\
=&&Y\mu_0dt + Y\sigma_0d\omega(t),
\end{eqnarray}
where both $\mu_0$ and $\sigma_0$ are time independent constants.
We further assume that $\mu_0$ and $\sigma_0$ are related by the 
equation $\mu_0=r+(\sigma_0'a^\star G)=(a^\star)'\alpha +
(a^\star)'(GG')a^\star(-1)+(\sigma_0'a^\star G)$. 
For any contingent claim $F=F(Y,t)$ having no explicit wealth 
dependence, the valuation partial differential equation Eq.~(\ref{eq:value1})
will simplify considerably:
\begin{equation}
{1\over 2} [tr(\sigma_0'\sigma_0)] Y^2 F_{YY} + rF_Y Y +F_t=rF,
\label{eq:bs1} 
\end{equation}
which holds for the economy which has a random production process and
random technological change as described above.

Suppose that the underlying stock for the contingent claim
follows a geometric Brownian motion
\begin{eqnarray}  
dS=&&S\mu_0dt+S\sigma_0 d\omega(t)\nonumber\\
  =&&S \mu_0 dt +S\sigma dZ(t),
\end{eqnarray}
where $\sigma^2=tr(\sigma_0'\sigma_0)$ and $Z(t)$ is a one dimensional
standard Wiener process. With this, the partial differential equation 
for the contingent claim $F(Y,t)=F(S,t)$ can be rewritten in the following
way:
\begin{equation}
{1\over 2} \sigma^2 S^2 F_{SS} + rF_S S +F_t=rF, 
\end{equation}
which is just Black-Scholes partial differential equation. Therefore, 
we have use general equilibrium point of the economy including random
production processes and uncertain technological change to derive
the Black-Scholes partial differential equation for contingent claim
pricing. Boundary conditions on various contingent claims may be imposed
to solve the above partial differential equation.  

\section{Summary}

In summary, I have considered an economy which includes uncertain production process
and the random technological change.  
When the uncertain production process has constant drift and 
variance, and the random technological change follows a 
stochastic log-normal process, it is shown that the general equilibrium
of such economy will lead to Black-Scholes partial differential 
equation satisfied by option prices. With slightest modification,
one could generalize this general equilibrium of random production
and uncertain technology economy to derive various partial
differential equations for option pricing when the underlying 
stocks have stochastic volatility. Such generalizations will be 
quite straightforward.  

\section{Acknowledgement}

I am indebted
to Professors P. Boyle and D. McLeish for introducing me to the field of  
modern finance. Encouragements from Dr. Y. Zhao, and my former
physics classmates of Princeton University Dr. R. Khuri 
and Dr.James T. Liu, are gratefully acknowledged. I am also
grateful to Dr.Daiwai Li, Dr.Jacqueline
Faridani, Dr.Chonghui Liu, Dr.Craig Liu, Dr. Z. Yang and Dr.Z. Jiang for interactions.
Any errors in this article are solely due to myself.

$^\dagger$ Email address: d6wang@barrow.uwaterloo.ca. Current address
is Toronto Dominion Bank. This paper
has been submitted to Journal of Financial Studies, Chinese Finance
Association, for publication.


\begin{references}
\bibitem{wang} Such scaling argument can easily apply to the case
where the van-Neumann-Morganstern utility function is power utility function.
\end{references}
\end{document}